\documentclass[prl,,twocolumn,amsmath,amssymb,floatfix]{revtex4}
\usepackage{graphicx}
\usepackage{bm}

\newcommand{\be}{\begin{equation}}
\newcommand{\ee}{\end{equation}}
\newcommand{\beq}{\begin{eqnarray}}
\newcommand{\eeq}{\end{eqnarray}}

\begin{document}

\title{Unzipping flux lines from extended
defects in type-II superconductors. }
\author{Yariv Kafri$^{1,2}$, David R. Nelson$^1$, and Anatoli Polkovnikov$^1$}
\affiliation{ {$^1$\small Department of Physics, Harvard University,
Cambridge, MA
  02138}\\
{$^2$\small Physicochimie Curie (CNRS-UMR168), Institut Curie,
Section de Recherche, 26 rue d'Ulm 75248 Paris Cedex 05 France}}

\date{\today}

\begin{abstract}
With magnetic force microscopy in mind, we study the unbinding
transition of individual flux lines from extended defects like
columnar pins and twin planes in type II superconductors. In the
presence of point disorder, the transition is universal with an
exponent which depends only on the dimensionality of the extended
defect. We also consider the unbinding transition of a single vortex
line from a twin plane occupied by other vortices. We show that the
critical properties of this transition depend strongly on the
Luttinger liquid parameter which describes the long distance physics
of the two-dimensional flux line array.
\end{abstract}

\maketitle

The competition between thermal fluctuations, pinning and
interactions of vortices in type-II high-temperature superconductors
leads to many interesting physical phenomena \cite{blatter}. These
include the melting of the Abrikosov flux-lattice into an entangled
vortex-liquid \cite{ns} and the proposed existence of low
temperature Bose-glass \cite{nv}, vortex glass \cite{fisher_glass}
and Bragg glass \cite{nat_sch} phases.

Experimental probes range from decoration, transport and
magnetization measurements, neutron scattering, electron microscopy,
electron holography to Hall probe microscopes. While these
experiments yield a wealth of information, the possibility of
manipulating a single vortex, for example using magnetic force
microscopy (MFM)~\cite{Wadas92}, has received less attention. Such
experiments could give rise to a direct measurement of microscopic
vortex properties which, up to now, have been under debate or
assumed. The possibility of performing such experiments is akin to
single molecule experiments on motor proteins, DNA, and RNA which
have transformed biophysics during the past decade by opening a
window on phenomena inaccessible via traditional bulk experiments
\cite{singlemol}.

Such an experiment was recently proposed by Olson-Reichhardt and
Hastings \cite{ORH04}. These authors suggested using MFM to wind two
vortices around each other, thus probing directly the energetic
barrier for two vortices to cut through each other. A high barrier
for flux line crossing has important consequences for the dynamics
of the entangled vortex phase.

In this Letter we analyze theoretically several possible experiments
in which a single vortex is depinned from extended defects using,
for example, MFM. We first consider the situation where MFM is used
to pull an isolated vortex bound to common extended defects such as
a columnar pin or a twin plane in the presence of point disorder.
Using a scaling argument we derive the displacement of the vortex as
a function of the force exerted by the MFM near the depinning
transition in an arbitrary dimension $d$. We argue that the
transition can be characterized by a universal critical exponent,
which depends {\it only on the dimensionality of the defect}. Hence
unzipping experiments from a twin plane directly measure the
free-energy fluctuations of a vortex in the presence of point
disorder in $d=1+1$ dimensions. To the best of our knowledge, there
is only one, indirect, measurement of this important quantity in
\cite{Bolle}. Our conclusions are supported by numerical and
analytical calculations. Related results apply when a tilted
magnetic field \cite{hatano} is used to tear away vortex lines in
the presence of point disorder.
\begin{figure}
\includegraphics[width=7.5cm]{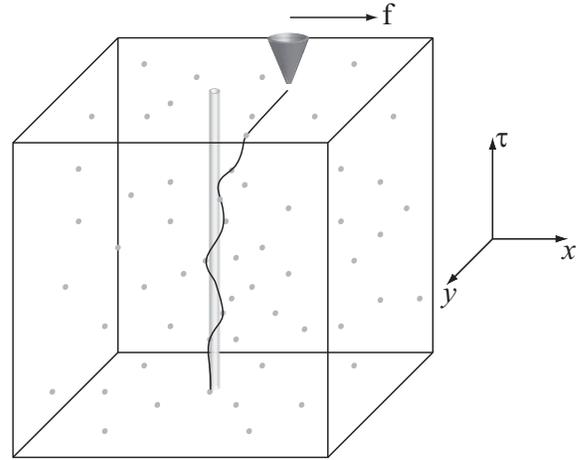}
\caption{The experimental setup considered in the text with a
columnar defect in 2+1 dimensions. The MFM is applying a force in
the $\hat{x}$ direction.}\label{singlevortex}
\end{figure}

Next, we study the effect of {\it interactions} between vortices
on the unzipping of a single vortex. We consider a system in which
vortices are preferentially bound to a thin two dimensional slab,
such as a twin plane, of say YBCO \cite{blatter}, in a three
dimensional sample. A similar situation can be achieved by
artificially inserting (with, for example, molecular beam epitaxy)
a thin plane with a reduced lower critical magnetic field $H_{c1}$
into a bulk superconductor; we require that the density of
vortices in the twin plane is much higher than in the bulk of the
sample. We imagine that MFM is used to pull a single vortex out of
the slab, thus creating an effective magnetic monopole inside it.
The displacement of the vortex from the slab as the transverse
force exerted by the MFM is increased then depends on the physics
of the two dimensional vortex liquid which resides in the slab.
Specifically, the ``Luttinger liquid parameter'' \cite{AHNS} which
controls large-distance behavior of the vortex liquid also
controls critical properties of the unbinding transition of the
vortex from the slab. We study the ``unzipping of Luttinger
liquids'' both with and without point disorder, and argue that
this setup can be used as a sensitive probe of the two dimensional
physics.

Consider first the unzipping of a single vortex from an extended
defect. We describe the flux line by a function ${\bf r}(\tau)$,
where $\tau$ denotes a coordinate parallel to the defect (see Fig.
\ref{singlevortex}), and ${\bf r}$ denotes the $(d-1)$-dimensional
coordinates perpendicular to it. In the absence of external force
the appropriate free energy for a given configuration ${\bf
r}(\tau)$ of the vortex is given by \cite{blatter}:
\begin{equation}
{\cal F}_0\!\!=\!\!\!\int_0^L \!\!\!\!d \tau \left[ \frac{\gamma}{2}
(\partial_\tau {\bf r}(\tau))^2+V({\bf r}(\tau))+\mu({\bf
r}(\tau),\tau) \right] .
\label{f0}
\end{equation}
Here $\gamma$ is the line tension and $L$ is the length of the
sample along the $\tau$ direction. The point disorder $\mu({\bf
r},\tau)$ is assumed uncorrelated and Gaussian distributed with
$\overline{\mu}=0$ and $\overline{\mu({\bf r},\tau)\mu({\bf
r'},\tau')}=\sigma \delta({\bf r}-{\bf r'})\delta(\tau-\tau')$,
where the overbar denotes an average over realizations of the
disorder. $V({\bf r})$ is a potential describing the
$d'$-dimensional extended defect.

To study unzipping, we consider a total free energy ${\cal F}={\cal
F}_0+{\cal F}_1$, where ${\cal F}_1=-{\bf f}\cdot {\bf
r(L)}=-\int_0^L {\bf f}\cdot
\partial_\tau \bf r(\tau)\,d\tau$. Here ${\bf f}$ stands for the
local force exerted by the MFM in the transverse direction and we
assume that ${\bf r}(0)=0$: below the unzipping transition the flux
line is unaffected by the boundary conditions at the far end of the
sample.

As the force, ${\bf f}$, increases the free energy density of the
unzipped portion of the vortex decreases. In contrast, the free
energy density of the bound part is, clearly, independent of the
force. The vortex will be unzipped when $f=f_c$ such that the
free-energy densities of the bound and unzipped states intersect. We
study the behavior of $s=\overline{\langle m \rangle}$, where $m$ is
the length along the $\tau$ direction which is unbound as the force
approaches the critical value $f_c$. The angular brackets and
overbar denote thermal and disorder averages, respectively. Although
we focus on disorder averages, particular disorder configurations
will be characterized by a pattern of jumps and pauses in the
unzipping~\cite{Lubensky}. Disorder averages can be probed
experimentally by unzipping many different vortices. It is
straightforward to show that $\langle m \rangle$ is related linearly
to the transverse displacement of the vortex at the top of the
sample, as measured by MFM. We take the force to always act in the
$x$ direction. Without disorder it is known that $s \sim
(f_c-f)^{-1}$ in any dimension and for any dimensionality of the
defect \cite{hatano}.

The universal properties of the unzipping transition with point
disorder can be analyzed using a simple scaling argument adapted
from~\cite{Lubensky} for the unzipping of DNA. Consider the free
energy ${\cal F}(m)$, with contributions from three sources. The
first is linear in $m$ and is due to the average free energy
difference between a vortex on the defect and in the bulk of the
sample $(f_c-f)m$. In addition there is a contribution, $\delta
{\cal F}_{d'}$, from optimized free-energy fluctuations of the
segment of the vortex which is bound to the defect. This term arises
from point disorder which is localized on or near the defect and is
expected to behave as $m^{1/2}$ at large $m$ for a $d'=1$
dimensional defect. For $d'>1$ we expect $\delta {\cal
F}_{d'}\propto m^{\omega(d')}$, where $\omega(d')$ is the exponent
associated with the free-energy fluctuations of a directed path in
the presence of point disorder in $d'$ dimensions \cite{Karreview}.
Finally, there is the interaction of the unzipped vortex with bulk
point disorder, $\delta {\cal F}_{d}\propto m^{\omega(d)}$, where
$d>d'$ is the dimensionality of the sample. Collecting these terms
gives:
\begin{equation}
{\cal F}(m)=a(f_c-f)m-b m^{\omega(d')}-c m^{ \omega(d) }\;.
\label{f}
\end{equation}
Here $a$, $b$ and $c$ are positive constants and the negative signs
have been chosen since the behavior is expected to be dominated by
the minima of the two random potentials. The exponent $\omega(d)$
has been studied extensively in the past and it is well known that
$\omega(d')<\omega(d)$ for any $d'<d$ \cite{Karreview}. For example
$\omega(d'=2)=1/3$ (twin planes) and $\omega(d=3)\simeq 0.22$ (bulk
sample). Therefore, disorder on or close to the defect controls the
unbinding transition for {\it any dimension} and the problem is
equivalent to unzipping from a sample with disorder localized on the
defect. In practice, disorder is likely to be concentrated within
real twin planes and near columnar damage tracks created by heavy
ion irradiation, strengthening even more the conclusions of this
simple argument. By minimizing the free energy with respect to $m$
we find the typical value of length of the unzipped part of the flux
line, which determines $s=\overline{\langle m \rangle}$:
\begin{equation}
s \sim \frac{1}{(f_c-f)^{\nu}}, \;\; \nu=[1-\omega(d')]^{-1}
 \label{nu}
\end{equation}
Thus we get $\nu=2$ (in agreement with~\cite{Lubensky}) for a
columnar pin and a new result, $\nu=3/2$ for a twin plane.

The heuristic argument leading to Eq.~(\ref{nu}) can be tested in a
number of ways. First consider unzipping from an attractive {\it
hard wall} in $(1+1)$-dimensions with point disorder in the bulk.
Upon applying an imaginary-gauge-transformation~\cite{hatano}
(possible below the unzipping transition) to the Bethe-ansatz
solution for $f=0$, obtained by Kardar (see Eq.~(3.12) in
Ref.~[\onlinecite{Kardar}]), we obtain
\begin{equation}
s=\int_0^{\infty} dy\, \kappa y \frac{\exp [ -(\lambda-f/  \gamma-
\kappa)y]}{[1-\exp(-\kappa y)]} \;. \label{kardar}
\end{equation}
Here $\kappa={\sigma/2}$ and $\lambda$ is the inverse localization
length characterizing the pinning to the wall in the absence of
point disorder. For $\kappa \ll \lambda-f/  \gamma$,
Eq.~(\ref{kardar}) reduces to the clean result: $s \sim (\lambda-f/
 \gamma)^{-1}$, while in the opposite limit one has $s \sim
(\lambda - \kappa - f/  \gamma)^{-2}$ as expected from Eq. \ref{nu}.
This result confirms our expectation that, even though disorder is
present everywhere, the universal properties of the unzipping
transition are determined by an effective disorder potential
generated in the vicinity of the wall with $\nu=2$.

\begin{figure}
\includegraphics[width=8.5cm]{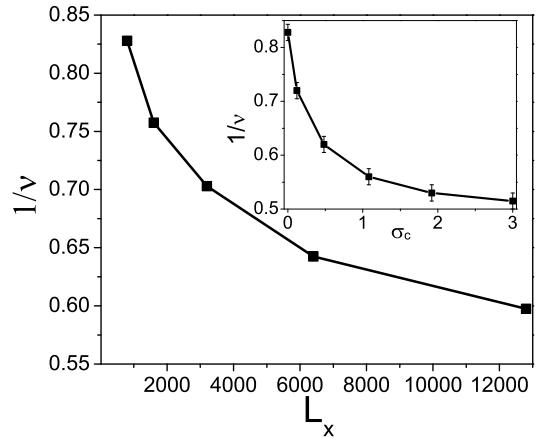}
\caption{Effective exponent $1/\nu$ versus $L_x$ for a fixed
strength of point disorder $\sigma=0.03$. The results are consistent
with the general argument that this exponent should saturate at
$\nu=2$ as $L_x\to\infty$. The inset shows the same exponent vs
$\sigma_c$, the variance of additional point disorder placed
directly on the columnar pin extracted from two system sizes
$L_x=600$ and $L_x=1200$. It is clear that $\nu\to 2$ as $\sigma_c$
increases.}
\label{fig1}
\end{figure}

To check the case of a symmetric attractive potential, representing
a columnar pin, we performed numerical simulations for a lattice
model in $d=1+1$. The restricted partition function of this model,
$Z(x,\tau)$, which sums over the weights of all path leading to
$x,\tau$, satisfies the recursion relation
\begin{eqnarray}
&&Z(x,\tau+1)=\delta_{x,0}(e^{V}-1)Z(0,\tau) \nonumber
\\ &&+e^{\mu(x,\tau+1)}\left[J
e^f Z(x-1,\tau)+J e^{-f}Z(x+1,\tau)\right].
\label{eqz}
\end{eqnarray}
Here $\mu(x,\tau)$ was taken to be distributed uniformly in
$[-U_0,U_0]$ with a variance $\sigma=U_0^2/3$. The variable $J$
controls the line tension ($J=\exp{\gamma/2}$) and was set to be
$J=0.1$ while we used $V=0.1$. We work in units such that $k_B T=1$,
where $k_B$ is the Boltzmann constant. The partition function was
evaluated for each variance of the disorder for several systems of
finite width $w=2L_x$ averaging over the time-like direction
(typically $\tau \simeq 10^6$ ``time'' steps) with the initial
condition $Z(0,0)=1$ and $Z(x,0)=0$ for $x \neq 0$. In the vicinity
of the transition we expect the finite size scaling prediction
$s=L_x f\left[L_x(f_c-f)^\nu\right]$. For finite $L_x$ we expect a
smooth interpolation between $\nu=1$ (the clean result) and $\nu=2$
(the asymptotic result with disorder) with increasing $L_x$ or
increasing strength of disorder. Using standard methods of
finite-size scaling we extracted the exponent for a given strength
of the disorder as a function of $L_x$. The exponent for each value
of $L_x$ is obtained from the best collapse of the data of two
systems sizes $L_x$ and $L_x/2$. As shown in Fig.~\ref{fig1}, the
data is consistent with $\nu$ saturating at $\nu=2$ for large
systems. The crossover to $\nu=2$ is much more rapid if the point
disorder is enhanced near the columnar pin (see the inset in
Fig.~\ref{fig1}), as might be expected for damage tracks created by
heavy ion radiation.

\begin{figure}
\includegraphics[width=8.5cm]{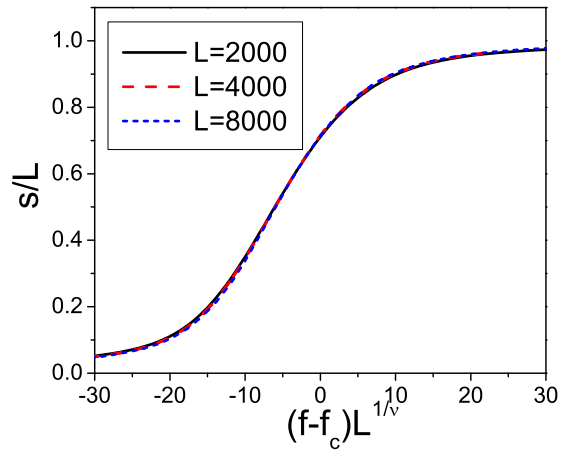}
\caption{Data collapse for unzipping from a disordered twin-plane
with $\nu=3/2$. See the text for more details.} \label{fig3}
\end{figure}

Numerics in $d=2+1$ dimensions with a twin plane or a columnar pin
proved more difficult. However, given that the universal properties
of the unbinding transition are determined only by the disorder on
the defect, we have performed numerical simulations of the model
with no bulk disorder and ignoring excursions of the vortex into the
bulk for $\tau$ values where it is pinned to the defect. This allows
very large systems to be studied. The $d=1+1$ problem, with no bulk
disorder, was studied in the context of DNA unzipping which yields
$s\propto (f_c-f)^{-2}$ in agreement with Eq.
\ref{nu}~\cite{Lubensky}. For additional analytic results, using the
replica trick see~\cite{knp}. In Fig.~\ref{fig3} we show a data
collapse with the anticipated exponent $\nu=3/2$ for unzipping from
a disordered twin plane ($d=2+1$), using the finite size scaling
form $s=L g \left[L(f_c-f)^\nu\right]$, with $L$ the size of the
system in the $\tau$ direction.
For the simulations we used a model like that in Eq.~(\ref{eqz})
with an in-plane hopping amplitude $J=1/2$ and a strength of point
disorder, $\sigma=4/3$. To obtain Fig.~\ref{fig3} we evaluated
numerically the free energy of the bound part of the vortex line for
each realization of point disorder and used an analytical expression
for the free energy of the unbound part ($F_u(m)$). The latter is
readily obtained from Eq.~(\ref{singlevortex}) by summing over
weights of paths of a directed random walk which is pulled by a
force $f$: $F_u(m)= -  f^2 m/ 2\gamma$. This allows a very efficient
calculation of $s$ including averages over point disorder. The
results agree very well with the prediction of Eq.~(\ref{nu}).

\begin{figure}[ht]
\includegraphics[width=7cm]{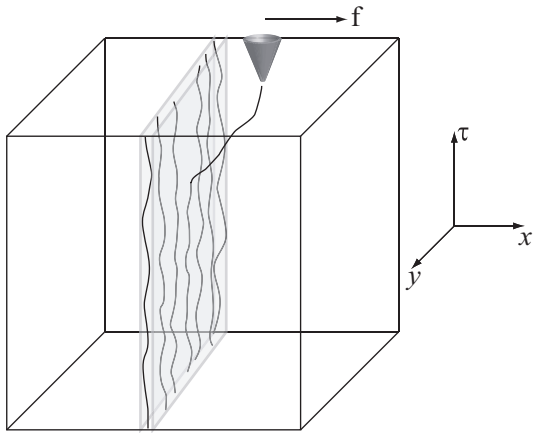}
\caption{Schematic representation of an MFM tip pulling out a vortex
out of a 1+1 dimensional flux line array.} \label{figLut}
\end{figure}

Consider now unzipping from a two dimensional plane with {\em many}
vortices in a three dimensional sample which is essentially free of
vortices (see Fig. \ref{figLut}). A MFM tip pulls the top end of one
of the vortices out of the plane with a force ${\bf f}=f \hat{x}$.
As in the single vortex problem, we expect an unzipping transition
for $f>f_c$ such that the vortex is pulled out of the two
dimensional slab.
We again write the free energy of the vortex as a sum of two
contributions: $F_u(m)$, which arises from the vortex segment
separated from the two dimensional slab, and $F_0(m)$, the free
energy change of the vortices which remain inside. Here $m$ is again
the length along the $\tau$ direction which is unbound from the
plane.

The free energy $F_u(m)$ is evaluated analytically as discussed
above. As for $F_0(m)$, clearly, there is a linear contribution
associated with the length $m$ removed from the attractive potential
of the slab. In addition, a {\it non-linear} contribution in $\tau$
arises due to the dislocation created in the vortex array (see Fig.
\ref{figLut}). The energy of this dislocation is determined by the
long distance properties of a two-dimensional vortex liquid. These
are known to be controlled by a single parameter $g$, which is
related to the elastic moduli of the vortex lattice and
temperature~\cite{AHNS}. For example, vortex density oscillations
without point disorder decay as a power-law in distance with an
exponent $\eta=2g$. Due to an analogy with one dimensional quantum
bosons, $g$ is often referred to as the ``Luttinger liquid
parameter''. Assuming that vortex lines must exit normal to the
interface, the energetic cost of the dislocation, $\mathcal F_d$,
can be calculated using the method of images: $\mathcal F_d={k_B
T\over 4g }\ln m$, for large $m$ \cite{chakin}.

Having obtained $F_0(m)$ and $F_u(m)$ it is straightforward to
calculate $\langle m \rangle=s$, (related to the mean displacement
of the MFM tip by $x_m=fs/\gamma$) as a function of $f$ using the
free energy
\begin{equation}
F(m)=r (f_c-f)m+{k_B T\over 4g }\ln(m).
\end{equation}
where $r$ is a positive constant. At the transition point, $f=f_c$
and $e^{-F(m)/T}$ is a power law in $m$, so the results are
sensitive to the value of $g$. For $g>1/4$ we find the same
divergence as for a {\it clean} twin plane without interactions,
\begin{equation}
s\sim {T\over f_c-f}.
\end{equation}
In contrast, for $1/8<g<1/4$ one finds a continuously variable
exponent governing the transition
\begin{equation}
s \sim \left({T\over f_c-f}\right)^{8g-1\over 4g},
\end{equation}
Finally, for $g<1/8$ we find that $s$ remains finite as $f\to f_c$
and the unzipping is presumably discontinuous. Higher moments of $m$
will also be sensitive to the value of $g$.

Finally, we consider the case of unzipping from a plane with many
vortices in a three dimensional sample in the presence of point
disorder. Here we expect the results to be sensitive to the boundary
conditions imposed on the plane. When the boundary conditions are
such that the number of flux lines is conserved (e.g. in a
cylindrical geometry) we expect the free energy fluctuations to
behave as $\delta {\cal F}(m)\propto m^{1/2}$ due to the highly
constrained, almost one dimensional, behavior of the flux lines.
Thus, the universal exponent is expected to be $\nu=2$ for any value
of $g$. In slab geometries however, where flux lines can freely
enter from the sides of the plane, the single flux line contribution
to the free-energy fluctuations can be screened and the effective
fluctuations come only from the dislocation energy. Near the vortex
glass transition at $g=1$,~\cite{fisher_glass,knp,Cardy} the
calculation can be carried out using the replica approach. Although
a rigorous treatment requires the method of images with reflected
disorder, we do not expect this to affect the results and performed
the calculations with uncorrelated disorder. We find that while the
quenched average of the free energy is independent of the disorder,
its fluctuations grow as $(\ln(m))^{1/2}$ for $g>1$ and
$(1-g)\ln(m)$ for $g<1$. Thus, in contrast to the single vortex
problem and to the case when the number of vortices in the sample is
conserved, here {\it fluctuations in the bulk} of the sample are
dominant, growing as $m^{\omega(3)}$. Using $\omega(3)\simeq 0.22$,
this lead one to the prediction $s\sim (f_c-f)^{-1.28}$, allowing a
direct measurement of $\omega(3)$.

We are grateful for discussions with O.~Auslaender, N.~Koshnik, C.
Moler, Y. Nonomura, and E. Zeldov. The work was supported by the
National Science Foundation through Grant No. DMR-0231631 and the
Harvard Materials Research Laboratory via Grant No. DMR-0213805. Y.K
was also supported by the Human Frontiers Science Program.

\end{document}